\documentclass[lettersize,journal]{article}
\usepackage{amsmath,amsfonts}
\usepackage{adjustbox}
\usepackage[table]{xcolor}
\usepackage{algorithm}
\usepackage{array}
\usepackage[caption=false,font=normalsize,labelfont=sf,textfont=sf]{subfig}
\usepackage{textcomp}
\usepackage{stfloats}
\usepackage{url}
\usepackage{verbatim}
\usepackage[noadjust]{cite}
\usepackage{graphicx}
\usepackage[noend]{algpseudocode}
\usepackage{siunitx}
\usepackage{authblk}
\algnewcommand{\TRUE}{\textbf{True}}
\algnewcommand{\FALSE}{\textbf{False}}

\begin{document}
\title{Routing and wavelength assignment in hybrid networks with classical and quantum signals}
\author{Lidia Ruiz}
\author{Juan Carlos Garcia-Escartin}
\affil{Departamento de Teor\'ia de la Se\~nal y Comunicaciones e Ingenier\'ia Telem\'atica, Universidad de Valladolid.\newline Paseo Bel\'en 15, 47011 Valladolid, Spain.\newline E-mails: lruiper@ribera.tel.uva.es (Lidia Ruiz), juagar@tel.uva.es (Juan Carlos Garcia-Escartin) }

\maketitle

\begin{abstract}
Quantum Key Distribution has become a mature quantum technology that has outgrown dedicated links and is ready to be incorporated into the classical infrastructure. In this scenario with multiple potential nodes, it is crucial having efficient ways to allocate the network resources between all the potential users. We propose a simple method for routing and wavelength assignment in wavelength multiplexed networks in which classical and quantum channels coexist. The proposed heuristics take into account the specific requirements of quantum key distribution and focus on keeping at bay the contamination of the quantum channels by photons coming from the classical signals by non-linear processes, among others. These heuristics reduce the shared path between classical and quantum channels and improve the signal-to-noise ratio in the quantum channels, improving their quantum key rate. We compare the results to the usual classical RWA approach.
\end{abstract}

\section{Introduction. QKD Networks}
Quantum Key Distribution (QKD) is one of the most mature quantum technologies. By combining the transmission of quantum signals through an insecure channel and public discussion through an authenticated, but public, classical channel, QKD permits two parties to generate a secret random key. The properties of quantum measurement guarantee that any eavesdropping attempts can be detected during the classical discussion. There exist multiple QKD protocols \cite{BB84,Eke91,SARG04,Noh09,SBC09,ZLA23} with experimental demonstrations and commercial products \cite{KLH15,MGM15,BAL17, XMZ20}.

In this paper, we will concentrate on discrete variable QKD over fiber optical networks where the quantum signal is made of optical pulses with less than one photon per pulse on average. There are also many continuous variable QKD protocols \cite{Ral99,CLA01,LPF18} which are more robust to crosstalk from classical channels. We do not consider them explicitly, but the general design principle of the routing algorithm still holds for these protocols (with extended noise thresholds).

While the first QKD realizations used dedicated dark fiber, there is a growing integration with classical networks \cite{MNR20,SGU22,CZW22}. As the number of end users grows, using dedicated point-to-point links is clearly impractical. This connection problem has already been solved in current optical fiber networks and the usual solutions can be readily adapted if we allow for some provisions for the quantum channels. 

Most of the classical infrastructure can be used directly in QKD. The most important exception is the use of amplifiers, which only introduce noise. The same limitations that prevent eavesdroppers from recovering the signal constitute an obstacle to the legitimate users, which can only generate keys at moderate distances. 

The number of available fiber links in any communication networks is limited and the channel resources must be shared by all the potential users. A common option is using Wavelength Division Multiplexing, or WDM, where multiple optical channels are transmitted at different wavelengths over the same optical fiber.

We propose a resource allocation algorithm for optical fiber networks with WDM supporting both classical and quantum networks. We consider the particular requirements of QKD and the physical sources of error, mostly through photon noise in the quantum channels due to non-linear processes in the classical signals. We are not considering advanced time multiplexing \cite{CZY17} or solutions with multicore fibers \cite{BdLC19}. We aim to describe a networking scheme which can be easily deployed on current Software Defined Networks that use the concept of \emph{lightpath}: a dedicated complete set of fiber links and all the required resources to allow the transmission from the origin to the destination node without conversion to the electronic domain or changes of wavelength. 

After introducing the problems of routing and wavelength assignment in optical fiber networks and classical-quantum multiplexing in Sections \ref{RWA} and \ref{CQMux}, respectively, we state our network model and assumptions in Section \ref{Model}. We will comment previous work on routing in QKD networks coexisting with classical channels (Section \ref{Related}) and define the problem we have solved in our experiments (Section \ref{Statement}). The proposed heuristics are given in Section \ref{Heuristics}. We conclude showing the results of our simulations and discussing in which cases this approach can improve the performance of the QKD network in Sections \ref{Results} and \ref{Conclusions}.

\section{Routing and Wavelength Assignment}
\label{RWA}
Given a network topology and a set of connection demands, the routing and wavelength assignment (RWA) problem consists in finding a way to allocate the network resources that maximizes the number of successful established communication channels \cite{RSS09}. Usually, there is a limited set of possible wavelengths for each fiber link, limited by the available hardware including lasers, multiplexers and detectors at each network node. 

Choosing the optimal path is a complex problem which is known to be NP-complete for the usual formulation \cite{CGK92}. An exact optimal solution is not practical in a networks with many nodes. In most cases, the RWA problem is solved by means of heuristics that are able to find a solution to the problem in polynomial time. We consider a typical scenario in a software defined network (SDN) where we assign the channels in real time as we receive requests for node to node connections. In this setting, we do not have the time or the computing power needed for optimal resource allocation and we recourse to heuristic algorithms, which give good results in classical networks \cite{ZJM00}.

We will propose three new heuristics designed to take into account the particularities of the quantum channels and compare them to existing classical heuristics (see Section \ref{Heuristics}).

\section{Classical-Quantum Multiplexing}
\label{CQMux}
Classical and quantum systems must coexist in QKD networks. QKD protocols need classical communication for the public discussion phase, but also for control information. It is also common having an optical synchronization signal running alongside the quantum states in the QKD protocol and using the same fiber for secure classical communication once we have generated a long enough secret key. Apart from that, we would like to be able to establish QKD links on top of existing networks carrying classical channels. 

Wavelength separation as used in classical multiplexing is not enough for successful coexistence of classical and quantum channels. The quantum link is delicate, with signal levels below the single photon. There are many ways the classical signals can contaminate the quantum channel, including Rayleigh backscattering \cite{SZT04}, linear crosstalk from the spectral tails of classical channels that are not sufficiently isolated from the quantum band \cite{HP85} or non-linear effects generating photons in the quantum wavelength \cite{PTC09,VGT23}. 

Most of these effects can be neglected when considering crosstalk between classical channels, but even a few photons are enough to make the quantum channel useless. The limiting factor for the quantum channel are nonlinear effects, in particular, Raman scattering and four-wave-mixing, FWM, \cite{FXT14,PTC09,EWL10,GSB21}. The dominant source of contamination depends on the transmitted signal power in the classical channels and the chosen frequency bands for the quantum and classical signals.
 
There are multiple potential wavelength assignments for the classical and the quantum channels. While the C-band seems more natural for the quantum signal due to the smaller losses, in many proposed systems the quantum signal is sent in the O-band around 1310 nm and the classical signals in wavelengths around 1550 nm so that the efficiency of the nonlinear effects converting photons from the classical bands to the quantum channel is low \cite{NTR05}. This wavelength scheme is not the only choice, but it shows the tradeoffs involved in the design of hybrid quantum-classical networks.

We will introduce these effects when we design our heuristic and give a routing and wavelength assignment scheme that takes into account the specific constraints of QKD networks. 

In this work, we will not consider some solutions like advanced channel allocation methods that explicitly search for frequency bands with good isolation \cite{NSC18}, choosing adequate transmission powers and modulation formats \cite{MKN17} or using multicore fiber \cite{BdLC19}. 

Our focus is on general simple methods that can be used over current fiber networks with minimal changes. The design can also be combined with other general methods like temporal filtering (limiting the measurement window with precise synchronization so as to minimize the effect of stray photons) \cite{PDC12}.

\section{Network model. Assumptions.}
\label{Model}
We will consider a passive optical network with no amplifiers or wavelength conversion. End-to-end communication is done with a dedicated path at a fixed wavelength (a lightpath). We assume we have a QKD software-defined network (QKD SDN) \cite{AHH17,CZW19,MAB19} where there is a classical control plane which uses our algorithm to evaluate all the connection requests and allocates the necessary resources at each node. 

We suppose the network has relatively short links below a hundred kilometers due to limitations in the achievable secure distance for commercial QKD links. For a point-to-point QKD link without quantum repeaters \cite{BDC98}, there is a linear bound on the secret key rate. Roughly, for any lossy channel with a transmission efficiency $\eta$, the maximum key rate $R$ is proportional to $\eta$ as losses become larger \cite{PGB09,TGW14,PLO17}, which falls exponentially with distance for optical fiber. We will not consider newer protocols with three parties (instead of two nodes), which started with the twin field QKD protocol \cite{LYD18}. In these protocols, while there is still a distance limit, the rate is only proportional to the square root of the transmission $O(\sqrt{\eta})$ \cite{LL18,MZZ18}. 

For these networks covering limited distances, we assume classical communication is perfect (the classical channels do not degrade in a noticeable way). Our bounded network could be part of a larger network with classical channels entering the subnet. If amplification were needed in this case, we can consider adding EDFA bypasses \cite{NRM06,AHW15,CZW19b}. 

Our model is focused on the bottleneck caused by the reduction of the length due to the noise the classical channels introduce into the quantum lightpaths. We will differentiate between the classical and the quantum lightpaths, as they require different resources. 

The control plane will receive a set of connection demands (an origin node and a destination) that arrive at random times. The RWA algorithm checks the current state of the network and, if it is possible, assigns a dedicated lightpath to the requested channels.

We consider two sources of traffic:
\begin{itemize}
\item{QKD channel demands.} The control plane receives a connection demand for an origin and a destination that want to generate a common secret key. We associate four lightpaths to this connection. A quantum lightpath from origin to destination to send the quantum states, two classical lightpaths for bidirectional communication between the nodes for synchronization and authenticated public discussion and a classical data channel that can be used for encrypted communication with the generated key. This is compatible with many existing QKD protocols. 
\item{Classical channel demands.} On top of the QKD network, we will consider the possibility of establishing individual lightpaths for classical channels not related to the QKD part of the network.
\end{itemize}
 
The performance of the RWA algorithm is measured in terms of established connections. The \emph{blocking probability} is the ratio between the number of refused connections and total demands. In classical optical networks a connection is blocked whenever there are no available resources. We consider we have enough transmitters and receivers for all the output links in a node. A classical connection will block only if there are no available free wavelengths in any possible path between origin and destination nodes. This also applies to the quantum channels.

However, we add a second blocking cause for the quantum links: we consider a QKD lightpath can be established only if it allows the secure generation of a secret key, at least in principle. In QKD protocols security can only be proved up to certain level of transmission errors and imperfections. The Qubit Error Rate, or QBER, determines which fraction of the quantum signals is interpreted incorrectly at the destination node after public discussion. In a perfect setting, when the origin and the destination use the same basis for state preparation and measurement, they should share the same bits (QBER$=\!0$). In the security proofs for QKD \cite{PR22} all the imperfections (QBER$>\!0$) are attributed to eavesdroppers. The QBER is used as a canary that detects their presence. For QBERs above a certain threshold, which depends on the protocol, we cannot safely extract a secret key and the whole key generation procedure must abort. 

If the noise due to the classical signals introduces more errors than the protocol can tolerate, even if we have available wavelengths, we should not establish a connection. The key generation procedure is doomed to fail and, meanwhile, the connection will waste network resources. We capture this in a conservative threshold that will consider that quantum connections block for a certain threshold of the quantum signal-to-noise ratio QSNR for each lightpath.

\subsection{Quantum signal to noise ratio}
In general, the formulas that relate the value of the QBER to the secret key rate and the thresholds for secure key generation can be difficult to compute and depend on the QKD protocol. We will use as a substitute a simpler metric which is directly related to the crosstalk noise: the quantum signal-to-noise ratio QSNR. In most natural scenarios, there is a direct relationship between a smaller QSNR (more noise relative to the quantum signal) and a growing QBER (a higher number of errors during detection at the receiver) \cite{CTP09,MWZ18,SWS18}.

We define the QSNR as the ratio between the quantum signal level $S_q$ and the noise level $N$ (photons introduced by crosstalk and other imperfections). We will use a simplified version
\begin{equation}
\text {QSNR}=\frac{Sq}{N}=\frac{n_R}{N_f+N_c}
\end{equation}
with the estimated photon number $n_R$ at the receiver after losses, including the effects of limited quantum efficiency at the detector and photon losses caused by dispersion taking part of the signal out of the detection window. The total noise is collected in a sum with a fixed term $N_f$ for imperfections such as dark counts at the detector, thermal noise or external light and a term for crosstalk noise $N_c$ both expressed as a photon number. 

These values can be adapted to the network at hand (depending on the transmission powers, losses, etc.). In our examples we consider the quantum signal is transmitted in the O band and estimate the photon number at the receiver from the expected photon number at the origin node after suffering attenuation during fiber transmission. The RWA algorithm will take the network graph and, for each path, it will compute the losses for a typical SMF-28 optical fiber at the transmission wavelength. 

The noise term should be estimated from the classical signal power. Raman scattering and four-wave-mixing grow with the optical power and the length of interaction. For all the discussed RWA algorithms we choose an upper bound for the number of generated noise photons considering the effective interaction length of the classical and quantum lightpaths. We use as an approximation a fixed coefficient multiplied by the total length in which there is a classical signal transmitted alongside the quantum lightpath. If multiple signals are transmitted at different wavelengths, we approximate the total effect as the sum of all of them. The coefficient can be adapted depending on the particulars of the studied network. Different optical fibers present different nonlinear coefficients. For simplicity, we have also considered the same noise for copropagation and counterpropagation (classical channels in the same and opposite direction of the quantum signals). This can be refined by defining two terms in $N_c$, one for each direction.

\section{RWA in QKD networks. Related Work}
\label{Related}
\rowcolors{2}{gray!25}{white}
\begin{table}[h!]
\centering
\begin{adjustbox}{width=1.2\textwidth}
 \begin{tabular}{|c|c|c|c|c|} 
\rowcolor{gray!50}
 \hline
Work & Network & Algorithms & \begin{tabular}{c} Optimization \\ criteria\end{tabular}  & \begin{tabular}{c} Distinguishing \\features \end{tabular} \\
 \hline
Cao et al. \cite{CZY17} & \begin{tabular}{c}  WDM network \end{tabular}  & \begin{tabular}{c} ILP for RTWA\\Heuristic (SP-FF)   \end{tabular} &  \begin{tabular}{c}Minimum resources\\Security level \end{tabular}  & \begin{tabular}{c} Considers key \\update periods \end{tabular}\\

Ning et al. \cite{NZY17} & \begin{tabular}{c} WDM network \end{tabular}  & \begin{tabular}{c} Soft-reservation \\ RTWA \end{tabular} &  \begin{tabular}{c} Accommodate variable \\security requirements \end{tabular} & \begin{tabular}{c} Time slot re-allocation \\ SP-FF\end{tabular}\\

Cao et al. \cite{CYZ18}& \begin{tabular}{c} QKPC over WDM networks \end{tabular}  & \begin{tabular}{c} ILP \\ Joint path-and-link RTWA \end{tabular} &  \begin{tabular}{c} Maximize accommodated\\key pool requests   \end{tabular}  & \begin{tabular}{c}  Introduces a time\\schedule  \end{tabular}\\

Zhao et al. \cite{ZCW18} & \begin{tabular}{c} WDM networks \end{tabular}  & \begin{tabular}{c} RTWA with $k$ SP-FF \end{tabular} &  \begin{tabular}{c} Minimize time conflicts \end{tabular}  & \begin{tabular}{c} Key updating period \\Time-sliding window\end{tabular}\\

Dong et al. \cite{DKZ20} & \begin{tabular}{c} QKD in trusted\\relay networks \end{tabular}  & \begin{tabular}{c} Two RTWA methods \end{tabular} &  \begin{tabular}{c}Reduce key resource consumption\\ Maximize score of physical security \end{tabular}   & \begin{tabular}{c} Quantum node bypass\\Auxiliary graphs\end{tabular}\\

Sharma et al. \cite{SPB21}& \begin{tabular}{c} QKD in trusted\\relay networks \end{tabular}  & \begin{tabular}{c} $k$ SP-FF RTWA \end{tabular} &  \begin{tabular}{c}Minimize blocking probability\\Improve time-slot use \end{tabular}  &  \begin{tabular}{c}Use of scheduling  \end{tabular}\\

Yu et al. \cite{YLZ22}  & \begin{tabular}{c} QKD in partially-trusted\\relay networks \end{tabular}  & \begin{tabular}{c} RTWA with \\ Collaborative routing \end{tabular} &  \begin{tabular}{c}Minimize key consumption\\Maximize secret-key supplement \end{tabular}   & \begin{tabular}{c} Analyzes different \\network topologies \end{tabular}\\

Chen et al. \cite{CZZ22} & \begin{tabular}{c} QKD in trusted\\relay networks \end{tabular}  & \begin{tabular}{c} Application priority\\ ranking \end{tabular} &  \begin{tabular}{c}Service priority\\ Key update rate \end{tabular}  &  \begin{tabular}{c} Allows retries \\ Designed for high concurrency \end{tabular}\\

Bi et al. \cite{BMD23} & \begin{tabular}{c} QKD in trusted\\relay networks \end{tabular}  & \begin{tabular}{c} Dynamic routing \end{tabular} &  \begin{tabular}{c}High remaining key volume\\ Low blocking probability \end{tabular}  &  \begin{tabular}{c}  Adjustable-size \\quantum key pool \end{tabular}\\

Chen et al. \cite{CCCZ23} & \begin{tabular}{c} QKD in hybrid trusted\\ and semi-trusted\\relay networks \end{tabular}  & \begin{tabular}{c} Multi-path routing\\ with k-shortest paths \end{tabular} &  \begin{tabular}{c} Maximize residual\\key volume \end{tabular}  & \begin{tabular}{c} Greedy algorithms \end{tabular}\\

Wenning et al. \cite{WPF23}  & KMN over DWDM & \begin{tabular}{c}3 protocols based on\\ $k$-shortest paths\end{tabular} & \begin{tabular}{c}Number of hops\\ Secret key rate\\ Buffer usage \end{tabular} &  Multilayer design \\ 

Zhang et al. \cite{ZAG23} & \begin{tabular}{c} QKD in trusted\\relay networks \end{tabular}  & \begin{tabular}{c} Joint Routing, Channel, Key-rate\\ and Time-slot Assignment \end{tabular} &  \begin{tabular}{c}Service priority\\ Key update rate \end{tabular}  & \begin{tabular}{c} Optional:\\  Trusted relays\\ Optical bypasses \end{tabular}\\

\hline
  \end{tabular}
\end{adjustbox}

\caption{Summary of related work on RWA for QKD networks. Used acronyms: KMN: Key Management Network. DWDM: Dense Wavelength Division Multiplexing. RTWA: Routing, Time and Wavelength Assignment. ILP: Integer Linear Programming. SP-FF: Shortest path, first fit. QKPC: Quantum Key Pool Construction. \label{TableRelated}}
\end{table}

Several studies in the literature have looked into the resource problem allocation for QKD. Table \ref{TableRelated} shows a selection of the previous work.

In many of them, the authors considered trusted or semi-trusted relays in networks with chains of trust \cite{DKZ20,SPB21,YLZ22,CZZ22,BMD23,CCCZ23,ZAG23}. Their focus is on building Key Management Networks where a few nodes are trusted. Resource allocation in those networks tries to create secure connections with the minimum resources (for different criteria). Security is discussed at a higher level, with key scheduling strategies and other security features. In all these cases, wavelength selection is only a way of to allocate channels. In many proposals, the authors combine both time and wavelength multiplexing to accommodate as many channels as possible \cite{CZY17,NZY17,CYZ18,ZCW18,DKZ20,SPB21,YLZ22,ZAG23}. In that sense, the strategies are not quantum, but a way to do efficient resource allocation for all the involved channels. 

In most cases, all the channels are assumed to be in the C band. While certain provisions, like placing a large guard band between the quantum and the classical traffic channels, can reduce some of the non-linear effects \cite{PTCR09}, in general, sharing links with classical data channels will affect the performance of existing quantum channels both in the C and O bands \cite{GFS21}. 

In contrast to the high level routing algorithms based on security, there are schemes that focus solely on wavelength allocation, without routing, with involved strategies that require a fine model of the fiber under study. One way to assign the channels is writing down and solving an optimization problem similar to RWA \cite{BRS18}. 

We will take an intermediate route, keeping the network model as simple as possible while capturing the most important sources of contamination (shared links between classical and quantum channels).

\section{Problem Statement}
\label{Statement}
We consider an arbitrary physical topology of an optical network as a directed graph $G(N,L, W_T,W_Q)$ composed of N nodes and L optical fiber links. $W_T$ is the number of wavelengths in each link, while $W_Q$ is the number of reserved wavelengths for the quantum channel. We consider quantum requests $R(s,d)$ where $s$ is the source node and $d$ is the destination node. In the experiments, we consider each quantum request includes four channels that must be established: the quantum channel, two classical control channels (one from $s$ to $d$ and one from $d$ to $s$) and an additional classical channel for user data transmission. Moreover, we assume the network to be a small Software Defined QKD Network \cite{YWZB17} in which the distance between each pair of nodes is within a certain range, so that the intermediate nodes between source and destination do not need to act as Trusted Relay Nodes. We assume end-to-end distances no longer than \SI{40}{\kilo\meter}.  

In this scenario, multiplexing classical and quantum channels together can introduce in the latter undesired noise due to Raman Scattering or four-wave-mixing effects \cite{CZY17}. We propose to reduce the the noise introduced by classical channels with three heuristics that allocate resources to the classical channels in a way that minimizes the shared distance with already established quantum channels.

\section{Heuristics}
\label{Heuristics}
We propose different algorithms to solve the RWA problem for the classical and quantum channels. We assume that to serve a quantum request we need to establish a quantum channel, two classical channels between source and destination for full duplex communication and a traditional data channel, which can carry an encrypted signal using the generated keys. Each lightpath is routed independently and two lightpaths between the same origin and destination can follow different paths. 

Resources are allocated to the quantum channel using $k$-Shortest Paths (KSP) and First Fit (FF) \cite{ZJM00}, i.e., the algorithm computes the $k$ shortest paths between source and destination and then checks for available wavelengths at the shortest path, allocating the first wavelength available. If no free wavelengths are found, the algorithm repeats the process with the second shortest path, the third, and so on until it finds an available path and wavelength or blocks otherwise. For the classical channels, we use KSP and FF and three new strategies:\\
\begin{itemize}

\item{Minimum Quantum Distance Overlap (MQDO):} The algorithm computes all the possible routes between source and destination and marks all the links shared with established quantum channels. Then, it tries to allocate an available wavelength at the path with the lowest accumulated distance in all the shared links using first fit. If the allocation is not possible, the algorithm repeats the process with the other paths in increasing order of shared link distance with the quantum channels. If it is not able to allocate resources, the connection is blocked. The pseudocode for this algorithm is shown in Algorithm \ref{alg:mqdo}.

\item{Minimum Quantum-Classical Channel Overlap (MQCCO):} Similar to the previous algorithm, MQCCO computes all the possible paths between the source and the destination nodes and calculates the distance shared with all the classical channels traversing any of the fibers in the route. Each used wavelength counts towards the distance estimation (two channels in a link are added as twice the distance). The main objective of the algorithm is to allocate the route with the minimum shared distance with classical channels traversing any of the fibers. The wavelength allocation is performed using first fit. If the allocation is not possible, the algorithm checks the other candidate paths in an increasing order of shared distance. If it fails to allocate a path and a wavelength, the request is blocked. The pseudocode for this algorithm is shown in Algorithm \ref{alg:mqcco}.

\item{Quantum Totally Disjoint (QTD):} The algorithm looks for a path between source and destination with no shared links with established quantum channels and uses first fit to allocate a wavelength. If no path is found, or the path has not available spectral resources, the connection is blocked. The pseudocode for QTD is shown in Algorithm \ref{alg:qtd}.

\end{itemize}

\begin{algorithm}
	\caption{Minimum Quantum Distance Overlap}
	\label{alg:mqdo}
	\begin{algorithmic}[1]
	\Procedure{Minimum Quantum Distance Overlap}{$source$, $destination$, $numWavelengths$, $candidatePathList$, $channelStatus$, $allocatedQuantumPaths$, $edgeLengthMatrix$}
		
		\State $candidatePathListCopy \gets$ \Call{copy}{$candidatePathList$}
		\State \Call{sort}{$candidatePathListCopy$, \text{Distance Overlap}}
		\For{$path$ in $candidatePathListCopy$}
			\State $w \gets 0$
			\While{$(w < numWavelengths)$}
				\State $available \gets \TRUE$
				\For{$link \in path$}
					\If{$channelStatus[link][w] = 1$}
						\State $available \gets \FALSE$
					\EndIf
				\EndFor
				\If{$available$}
					\State \textbf{return} $path, w$
				\Else
					\State $w \gets w + 1$
				\EndIf
			\EndWhile
		\EndFor
		\State \textbf{return} $[], -1$
	\EndProcedure
	\end{algorithmic}
\end{algorithm}

\begin{algorithm}
	\caption{Distance Overlap (used in MQDO)}
	\begin{algorithmic}[1]
		\Procedure{Distance Overlap}{$path$}
			\State \textbf{global} $edgeLength$
			\State \textbf{global} $graphNetworkEdges$
			\State \textbf{global} $allocatedQuantumPaths$
			\State $length \gets 0$
		
			\For{$link \in path$}
				\For{$key$ in $allocatedQuantumPaths$}
					\For{$qPath$ in $allocatedQuantumPaths[key]$}
						\If{$link \in qPath$}
							\State $length \gets length + edgeLength[link]$
						\EndIf
					\EndFor
				\EndFor
			\EndFor
			\State \Call{print}{$length$}
			\State \textbf{return} $length$
		\EndProcedure
	\end{algorithmic}
	\label{alg:distance_overlap}
\end{algorithm}

\begin{algorithm}
\caption{Minimum Quantum-Classical Channel Overlap}
\begin{algorithmic}[1]
\Procedure{Minimum Quantum-Classical Channel Overlap}{$source$, $destination$, $numWavelengths$, $candidatePathList$, $channelStatus$, $allocatedQuantumPaths$}
		
		\State $sortedCandidatePathList \gets$ \Call{copy}{$candidatePathList$}
		\State \Call{sort}{$sortedCandidatePathList$, \text{Channel Overlap}}
		\For{$path$ in $sortedCandidatePathList$}
			\State $w \gets 0$
			\While{$(w < numWavelengths)$}
				\State $available \gets \TRUE$
				\For{$link \in path$}
					\If{$channelStatus[link][w] = 1$}
						\State $available \gets \FALSE $
					\EndIf
				\EndFor
				\If{$available$}
					\State \textbf{return} $path, w$
				\Else
					\State $w \gets w + 1$
				\EndIf
			\EndWhile
		\EndFor
		\State \textbf{return} $[], -1$
	\EndProcedure
\end{algorithmic}
\label{alg:mqcco}
\end{algorithm}

\begin{algorithm}
	\caption{Channel Overlap (used in MQCCO)}
	\begin{algorithmic}[1]
		\Procedure{Channel Overlap}{$path$}
			\State \textbf{global} $edgeLength$
			\State \textbf{global} $graphNetworkEdges$
			\State \textbf{global} $allocatedQuantumPaths$
			\State \textbf{global} $channelStatus$
			\State $length \gets 0$
		
			\For{$link$ in $path$}
				\State $occupiedWavelengths \gets$ \Call{sum}{$channelStatus[link]$}
				\For{$key$ in $allocatedQuantumPaths$}
					\For{$qPath$ in $allocatedQuantumPaths[key]$}
						\If{$link \in qPath$}
							\State $length \gets length + occupiedWavelengths \times edgeLength[link]$
						\EndIf
					\EndFor
				\EndFor
			\EndFor
			\State \textbf{return} $length$
		\EndProcedure
	\end{algorithmic}
	\label{alg:channel_overlap}
\end{algorithm}

\begin{algorithm}
	\caption{Quantum Totally Disjoint}
	\begin{algorithmic}[1]
		\Procedure{Quantum Totally Disjoint}{$source$, $destination$, $numWavelengths$, $candidatePathList$, $channelStatus$, $allocatedQuantumPaths$}
			\For{$path$ in $candidatePathList$}
				\State $search \gets \TRUE $
				\State $foundPath \gets \TRUE $ 
				\While{$search$}
					\For{$key$ in $allocatedQuantumPaths.keys()$}
						\For{$qPath$ in $quantum\_paths[key]$}
							\For{$link \in path$}
								\If{$link \in qPath$}
									\State $search \gets \FALSE$
									\State $foundPath \gets \FALSE $
									\State \textbf{break}
								\EndIf
							\EndFor
						\EndFor
					\EndFor
					\State $search \gets \FALSE$
				\EndWhile
				\If{$foundPath$}
					\State $w \gets 0$
					\While{$(w < numWavelengths)$}
						\State $available \gets \TRUE$ 
						\For{$link \in path$}
							\If{$channelStatus[link][w] = 1$}
								\State $available \gets \FALSE $
							\EndIf
						\EndFor
						\If{$available$}
							\State \textbf{return} $path, w$
						\Else
							\State $w \gets w + 1$
						\EndIf
					\EndWhile
				\EndIf
			\EndFor
		\State \textbf{return} $[], -1$
		\EndProcedure
	\end{algorithmic}
	\label{alg:qtd}
\end{algorithm}

If the algorithms are able to find resources, they check if the new path meets the requirements of QSNR for secure QKD at a reasonable key generation rate in the quantum channels. If the resulting QSNR is below a user defined QSNR threshold, the request is blocked. Likewise, it is checked if the found paths to be allocated to the classical channels would affect the QSNR of already established quantum channels. If the classical data channels lower any established quantum channel’s QSNR below a threshold, the request is also blocked. The QSNR is estimated with:

\begin{equation}
\text{QSNR}\ =\frac{e^{-\alpha L^q} P_{tx}}{N_{fiber}+\sum_{i} N_{shared}\cdot L_{i,\ shared}}\ 
\end{equation}
where $\alpha$ is the attenuation of the fiber, $L^q$ is the length of the route allocated to quantum channel $q$, $P_{tx}$ is the transmitted power, $N_{fiber}$ is a fixed noise term, $N_{shared}$ represents the noise introduced from one classical channel into the quantum channel per km and $L_{i,shared}^q$ is the link distance that the classical channel $i$ shares with the the quantum channel $q$. $P_{tx}$, $N_{fiber}$ and $N_{shared}$ are given as a photon number.

\section{Simulation Set up and Results}
\label{Results} 
\begin{figure}[!t]
\centering
\includegraphics[width=2.5in]{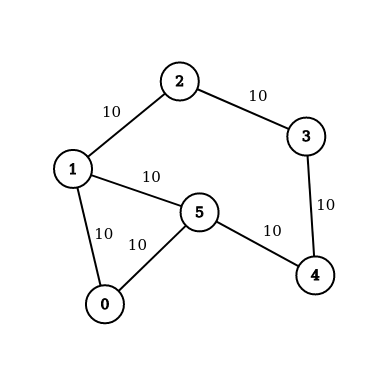}
\caption{Graph for the example network under study. We consider 6 nodes with 7 bidirectional links of 10 km each.}
\label{graph}
\end{figure}

We study the effects of the proposed heuristic algorithms on the QSNR of quantum channels in a small topology composed of 6 nodes 14 links, shown in Figure \ref{graph}. We assume 80 wavelengths, 40 for the quantum channel in the O-band and 40 for the classical channels in the C-band. We consider a QSNR threshold of $\SI{15}{\deci\bel}$ for secure QKD. We set the parameter $N_{fiber}= 3.8\cdot 10^{-4}$ and choose $N_{shared}$ so that $\text{QSNR} =\ 31.5$ ($\SI{15}{\deci\bel}$) when $L^q\ =\SI{40}{\kilo\meter}$ and $\sum_{i} L_{i,\ shared}^q\ =\SI{40}{\kilo\meter}$. We have selected a scenario where noise from the classical channel is relevant. A single classical channel does not interfere significantly with the quantum transmission, but the combined effect of multiple channels and longer paths might. We consider a fiber attenuation $ \alpha=\SI{0.32}{\deci\bel/\kilo\meter}$ at $\SI{1310}{\nano\meter}$ \cite{Corning} and a normalized $P_{tx}=1$. Lastly, we consider $k = 5$ possible paths when performing $k$-Shortest Paths.

We calculate the blocking ratio and the average QSNR for a number of total requests arriving at the networks ranging from $1$ to $100$ with an increment of $1$ requests in the range from $1$ to $10$ and then in steps of $10$ requests from $10$ to $100$. We consider static traffic requests between uniformly chosen random origin and destination nodes. For each number of requests, $1500$ simulations were run. Results are represented with $95\%$ confidence intervals.

\begin{figure}
\centering
\includegraphics[width=2.5in]{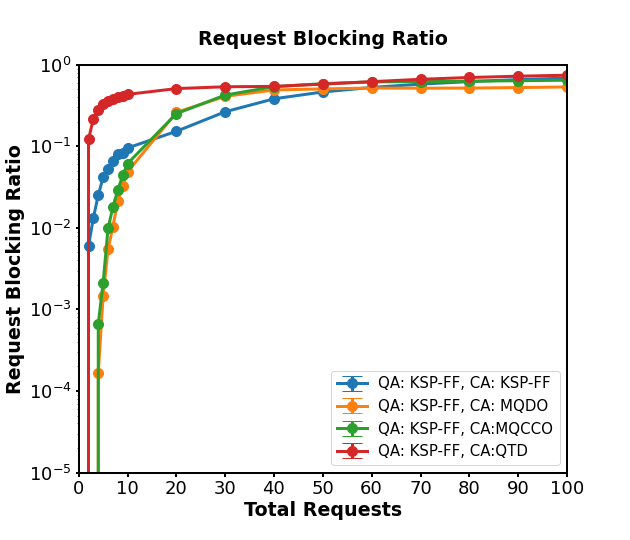}
\caption{Blocking ratio for the proposed heuristic algorithms for the classical channels. All the quantum channels are routed using KSP-FF. The error bars (95\% confidence) are too small to be noticed.}
\label{br}
\end{figure}

Figure \ref{br} shows the blocking ratio for the proposed algorithms and how they compare to the usual KSP-FF heuristic used in classical networks. Since QTD tries to allocate the classical channels in a completely disjoint path with respect to the quantum channels, resources are exhausted sooner compared to the other algorithms and, therefore, the blocking ratio is the highest. On the other hand, MQDO and MQCCO achieve very similar blocking ratios and improve the performance of KSP-FF up to an order of magnitude when the number of requests are lower than $15$. Since MQDO and MQCCO balance the network load between the quantum and the classical channels, they are able to establish more connections than KSP-FF. 

\begin{figure}[!t]
\centering
\includegraphics[width=2.5in]{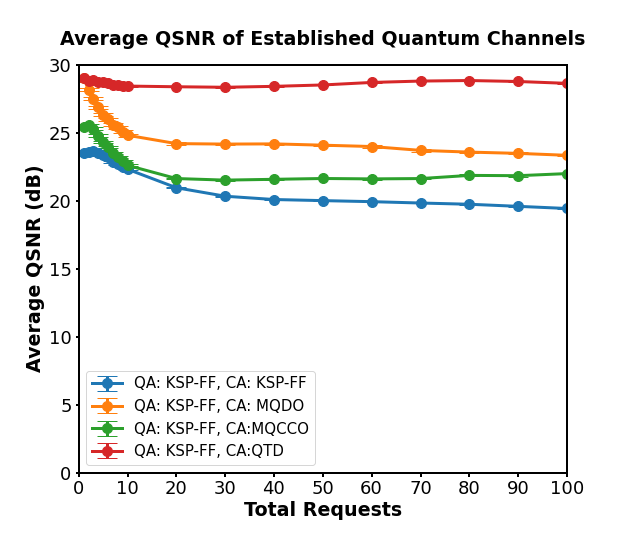}
\caption{QSNR of the served QKD requests using the KSP-FF, MQDO, MQCCO and QTD algorithms for classical channel allocation. All the quantum channels are routed using KSP-FF. The points show the average behaviour and the error bars show the variance for our random traffic requests.}
\label{snr}
\end{figure}

Figure \ref{snr} shows the average QSNR of the established quantum channels. It can be seen that QTD achieves the highest QSNR, i.e., the algorithm establishes less quantum channels than other QKD-specific resource allocation algorithms like MQDO and MQCCO, but the established quantum channels have the highest QSNR, up to $\SI{7}{\deci\bel}$ above the results of MQCCO and $\SI{5}{\deci\bel}$ above MQDO. This means that QTD achieves the quantum connections with the highest key rate. On the other hand, MQDO achieves a better QSNR than MQCCO (up to $\SI{4}{\deci\bel}$ in improvement), while obtaining similar blocking ratios. Including the number of classical channels that traverse the potentially shared fiber as a restriction when finding the quantum path does not lead to a better performance neither in  terms of blocking ratio nor in terms of QSNR and it leads to a slightly more complex algorithm. Lastly, KSP-FF, which was designed to allocate resources in classical networks, obtains the lowest QSNR, i.e., the worst quality quantum channels. In either case, although the blocking ratio of all the algorithms is high, the quality of the established quantum channels surpasses the QSNR threshold by a significant margin (by $\SI{5}{\deci\bel}$ for KSP-FF in the worst case, almost by $\SI{14}{\deci\bel}$ QTD in the best case). Blocking low QSNR requests seems a viable strategy, even if routing is done with traditional algorithms.

As explained in Section \ref{Heuristics}, the path returned by the QRWA algorithms undergoes a validation process. In the case of quantum channels, this process verifies that the resulting QSNR meets the QSNR requirements. Conversely, in the case of classical channels, the validation involves ensuring that the path does not compromise the established quantum channels' QSNR below the predefined threshold. This find a route first, check the threshold later strategy could lead to a higher blocking ratio, since the algorithms are prevented from exploring alternative routes that could meet with these requirements. This problem can be easily addressed by including an estimation of the QSNR and QSNR degradation of the candidate paths within the QRWA algorithms for quantum and classical channels respectively. This way, the QRWA algorithms discard the candidate path if it fails to meet the QSNR requirements and explore a new one until a path that meets the requirements and has available wavelengths is found or after all candidate paths are evaluated. 

\begin{figure}
	\centering
	\includegraphics[width=2.5in]{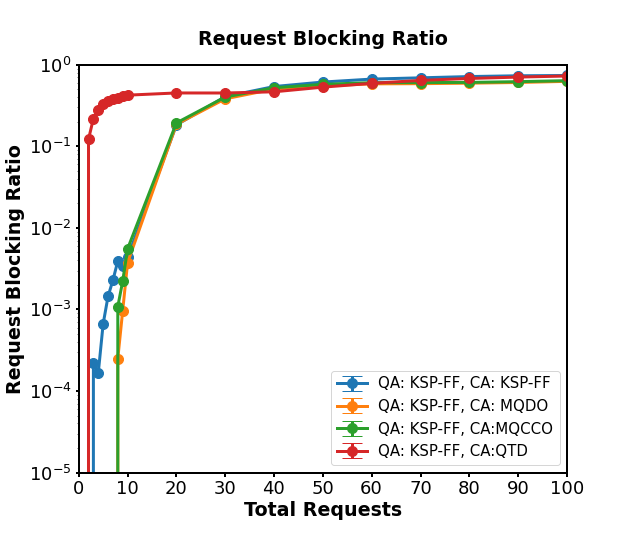}
	\caption{Blocking ratio for the proposed heuristic algorithms when the QSNR validation is performed within the QRWA algorithms. All the quantum channels are routed using KSP-FF.}
	\label{total_br_with_qsnr_estimation}
\end{figure}

\begin{figure}[!t]
	\centering
	\includegraphics[width=2.5in]{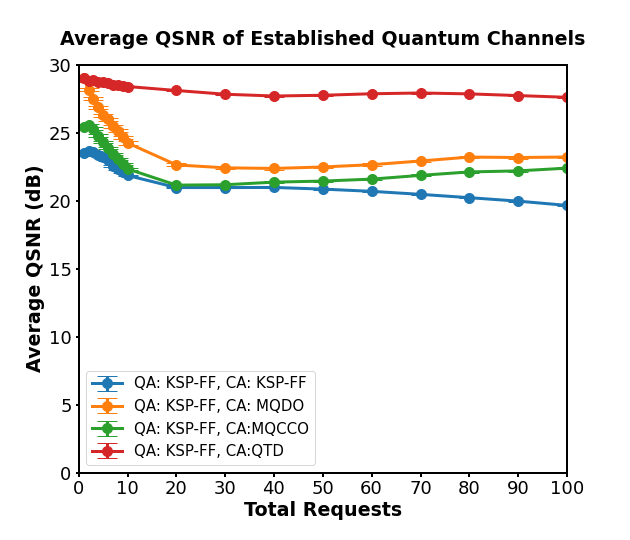}
	\caption{QSNR of the served QKD requests using the KSP-FF, MQDO, MQCCO and QTD algorithms for classical channel allocation when the QSNR validation is performed within the QRWA algorithms.}
	\label{total_snr_with_qsnr_estimation}
\end{figure}

Figures \ref{total_br_with_qsnr_estimation} and \ref{total_snr_with_qsnr_estimation} show the blocking ratio and the QSNR obtained when the QSNR and QSNR degradation requirements are evaluated within the algorithms. It can be seen that the blocking ratio improves for KSP-FF, MQDO and MQCCO when the total requests arriving to the network is lower than $10$. In particular, MQDO and MQCCO are able to establish all the channels for up to $8$ total requests, while the blocking ratio for KSP-FF lowers in up to two orders of magnitude. On the other hand, the QSNR suffers a slight penalty compared to the QSNR shown in Figure \ref{snr} for QTD and MQDO. However, the general behaviour of the QRWA algorithms does not change, i.e., QTD obtains the highest QSNR at the expense of blocking more requests and MQDO and MQCCO improve the performance of KSP-FF in terms of QSNR, establishing channels with higher quantum key rate.

\subsection{QKD over a classical network}

Let us study now what happens to the blocking ratio and the QSNR when the QKD network shares the infrastructure with purely classical channels. To this end, we will assume a maximum of $10$ requests, which is the maximum number of total requests for which MQDO and MQCCO presented a better blocking ratio than KSP-FF. We define the traffic classical load as the fraction of the total requests that are classical requests. This load varies in increments of $0.1$ from $0.0$ (no classical requests) to $1.0$ (all channels are classical).

The code used for our experiments can be found online \cite{Rui24}.

\begin{figure}[!t]
\centering
\includegraphics[width=2.5in]{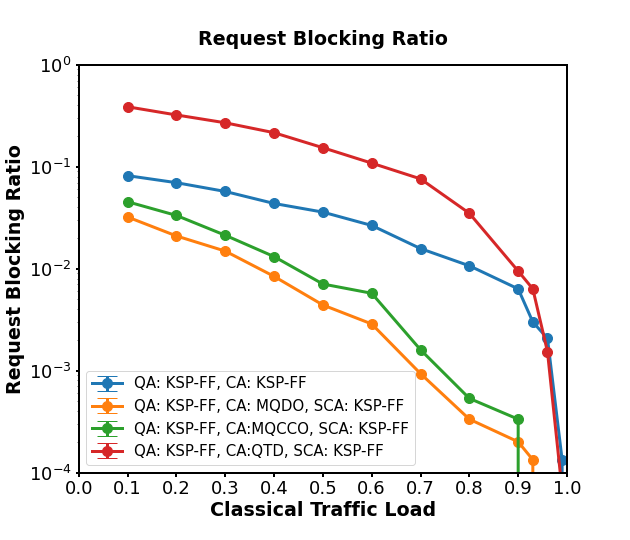}
\caption{Blocking ratio of the proposed algorithms when the network is shared with classical connections. Variance too small to be appreciable.}
\label{blocking_ratio_classical}
\end{figure}

Figure \ref{blocking_ratio_classical} shows the blocking ratio for all the algorithms. It can be seen that the blocking ratio decreases when the classical traffic load increases. Since a classical request requires just one lightpath to be established, it is easier to accommodate that single channel than a quantum request, which requires four lightpaths in our scenario (one quantum channel, two control channels and one data channel). Hence, increasing the classical load leads to a better performance of the network. However, this can correspond to a higher fraction of quantum requests that are blocked. 

\begin{figure}[!t]
\centering
\includegraphics[width=2.5in]{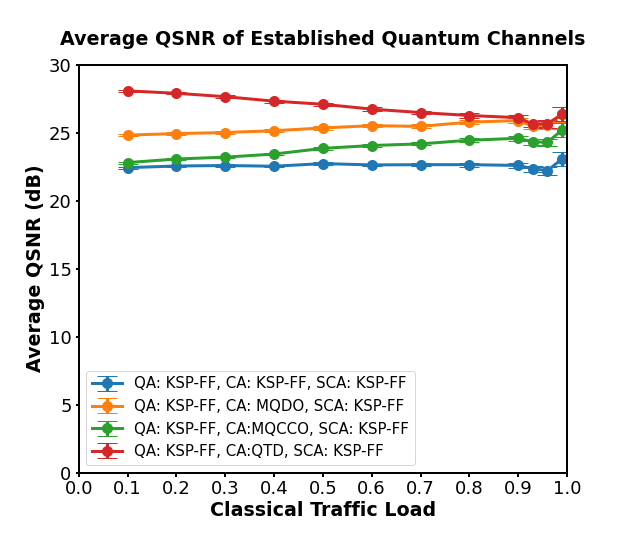}
\caption{QSNR of the served QKD requests when the network is shared with classical connections.}
\label{snr_classical}
\end{figure}

If we observe the evolution of the average QSNR for the established quantum channels in Figure \ref{snr_classical}, it clearly degrades for the QTD algorithm for all classical loads. The algorithm may have to find longer paths for the quantum channels for them to be totally disjoint from the classical paths, leading to a decrease of the QSNR. On the other hand, the MQDO and MQCCO, as well as the KSP-FF algorithms, get a fairly stable QSNR when the classical traffic load increases. Hence, although the connections solved using QTD are not able to maintain the quality of the quantum channels when the network is shared with classical requests, they still have a better QSNR than the served requests with the other proposed algorithms, up to $6\, \text{dB}$ above MQCCO and $3\, \text{dB}$ better than MQDO and KSP-FF. 

\section{Conclusions}
\label{Conclusions}
We have proposed three heuristic RWA algorithms that allocate resources to classical channels in QKD networks. The primary objective of these algorithms is minimizing the shared distance with quantum channels. This way, we can reduce undesired effects like Raman Scattering or four-wave-mixing. A comparative analysis against the well known classical RWA algorithm, Shortest-Path First Fit, shows an improvement of the blocking ratio up to one order of magnitude. Additionally, the QSNR can improve by up to $9\, \text{dB}$, leading to quantum connections with higher key rates. 

All the heuristics are easy to compute and similar to successful classical heuristics, but include enough detail to model the non-linear effects that are physically relevant to describe the noise the classical signals introduce into the quantum channels. 

The most strict approach of choosing independent routes for classical and quantum channels gives the best QSNR, but can lead to rejecting channel requests that could be accepted with a small performance penalty. Other QKD-specific resource allocation algorithms like MQDO and MQCCO seem a better solution. All the proposed approaches improve the QSNR respect to the classical algorithm KSP-FF, showing that using specific resource allocation methods for QKD channels improves the overall performance of hybrid quantum-classical networks.

\section{Acknowledgments}
Lidia Ruiz has been funded by the European Union NextGenerationEU (PRTRC17.I1) and the Consejer\'ia de Educaci\'on, Junta de Castilla y Le\'on, through QCAYLE project. J.C. Garcia-Escartin has been funded by the the European Union NextGenerationEU (PRTRC17.I1) and the Consejer\'ia de Educaci\'on, Junta de Castilla y Le\'on, through QCAYLE project and the Spanish Government and FEDER grant PID2020-119418GB-I00 (MICINN).

\newcommand{\noopsort}[1]{} \newcommand{\printfirst}[2]{#1}
  \newcommand{\singleletter}[1]{#1} \newcommand{\switchargs}[2]{#2#1}

\end{document}